# DESIGN OF 11.8 MHZ BUNCHER FOR ISAC AT TRIUMF

A.K. MITRA, R.L. POIRIER, R.E. LAXDAL, TRIUMF

Abstract

The high energy beam transport (HEBT) line for the ISAC radioactive beam facility at TRIUMF requires an 11.8 MHz buncher. The main requirements of the buncher are to operate in cw mode with a velocity acceptance of 2.2% and an effective voltage of 100 kV, which for a three gap buncher gives a drift tube voltage of 30 kV. A lumped element circuit is more suitable than a distributed rf structure for this low frequency of operation. The resonant frequency of 11.8 MHz is obtained by an inductive coil in parallel with the capacitance of the drift tube. The coil is housed in a dust free box at atmospheric pressure whereas the drift tube is placed in a vacuum chamber and an rf feedthrough connects them. Two design of this feedthrough, one using disk and one using tubular ceramics, operating at 30 kV rf, are described in this paper. MAFIA and SUPERFISH codes are used to simulate the fields in the feedthroughs, particularly around the ceramic metal interfaces. Test results of the prototype feedthroughs are presented and the choice of the proposed final solution is outlined.

## 1 INTRODUCTION

The beam from the DTL of the ISAC radioactive beam facility goes thru the high energy beam transport (HEBT) and is delivered to various target stations. The DTL produces beams fully variable in energy from 0.15-1.5 MeV/u with mass to charge values of $3 \leq A/q \leq 6$. A low-$\beta$ 11.78 MHz buncher placed approximately 12 m down stream from the DTL can provide efficient initial bunching for beams from 0.15-0.4 MeV/u [1]. The basic parameters of the HEBT low-$\beta$ buncher are given in Table 1.

Table 1: Basic parameters of the HEBT low-$\beta$ buncher

| Resonant frequency, f | 11.78 MHz |
|---|---|
| Velocity ($\beta c$) | 0.022 |
| Charge to mass ratio | $1/3 \geq q/A \geq 1/6$ |
| Energy range | 0.15 –0.4 MeV/u |
| Veffective, maximum | 100 kV |
| Vtube | 30 kV |
| Number of gaps | 3 |
| $\beta\lambda/2$ | 29.28 cm |
| Beam aperture, diameter | 2.0 cm |
| Cavity length | 70.0 cm |
| Voltage stability | ± 1.0% |
| Phase stability | ± 0.3 % |
| operation | cw |

The maximum required effective voltage from the buncher is 100 kV. For a 3 gap structure, the effective voltage, Veff is given by $V_{eff} = 4\, V_t\, T_o$, where $V_t$ is the drift tube voltage and $T_o$ is the transit time factor.

## 2 DESIGN

A prototype of a two gap structure is designed to produce 30 kV tube voltage at the HEBT buncher frequency. Since the resonant frequency is low, a lumped element circuit is found to be more suitable than a distributed structure. An inductive coil in parallel with the capacitance of the drift tube and circuit capacitances produces the desired resonant frequency. The coil, shorted at one end, is placed in a dust free box. Two designs of feedthroughs are tested. An rf feedthrough connects the open end of the coil to the drift tube, which is in a vacuum box. Two nose cones are also attached in the vacuum box to simulate the gap capacitance of the HEBT buncher. The prototype buncher is shown in Fig. 1. Since the power dissipated is estimated to be approximately 700 watts, the coil is water cooled. Insulators support the coil to reduce vibration due to water flow.

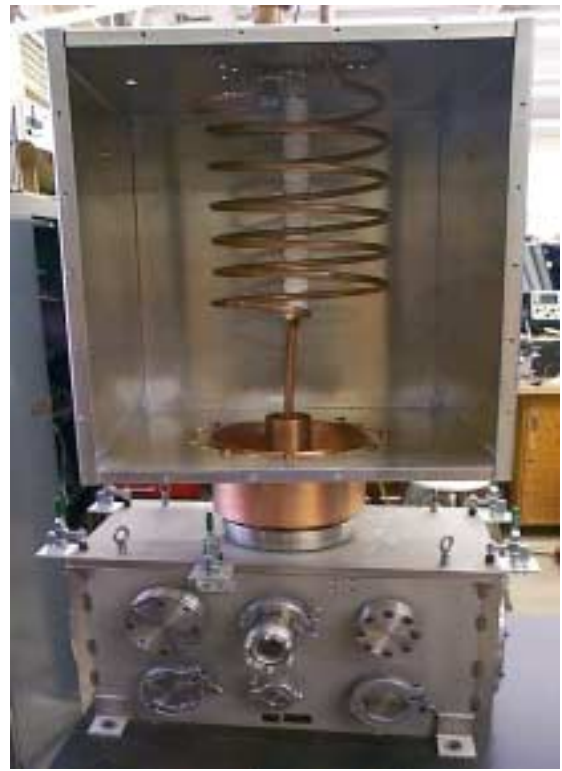

Figure 1: Prototype of HEBT buncher

## 2.1 The coil

A 6 turn coil is made of a hollow copper tube of ½" diameter. Inductance of this coil is 6.1 µH with turn-to-turn spacing of 1.5", coil diameter of 9" and a coil length of 15". A ¼" hollow tube is inserted in this ½" tube before the coil is made. Water flows through this ¼" tube and flows out from the ½" tube. The coil is installed in an aluminum box 20"x20"x24" and the water inlet, outlet and coupling loop are located on top of this box.

## 2.2 Disk ceramic feedthrough

The disk feedthrough [2] uses a ceramic disk of 4.5" outer diameter with inner hole of 1.75" diameter and a thickness of 0.375". The ceramic is not metalized and it is held in position by bolting two halves of both inner and outer conductors as shown in Fig. 2. Helicoflex rings are used for vacuum seal between ceramic and the metal parts. In case of window failure, only the ceramic need be replaced and the metal parts can be reused. The MAFIA static solver is used to design the contour of the metal around the ceramic and is shown in Fig. 3

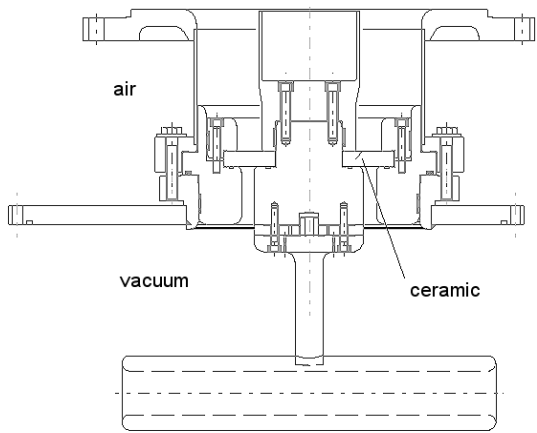

Figure 2: Sectional view of disk ceramic feedthrough

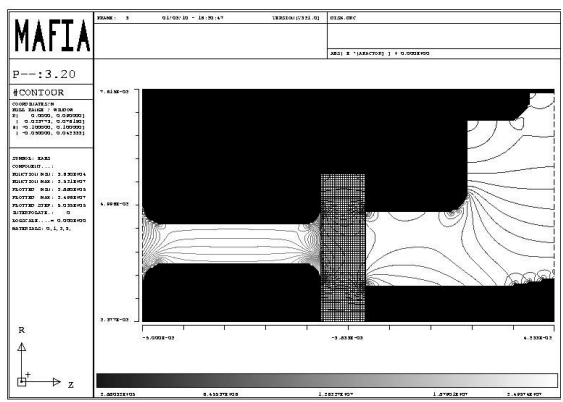

Figure 3: MAFIA plot of e-fields

## 2.3 Tubular ceramic feedthrough

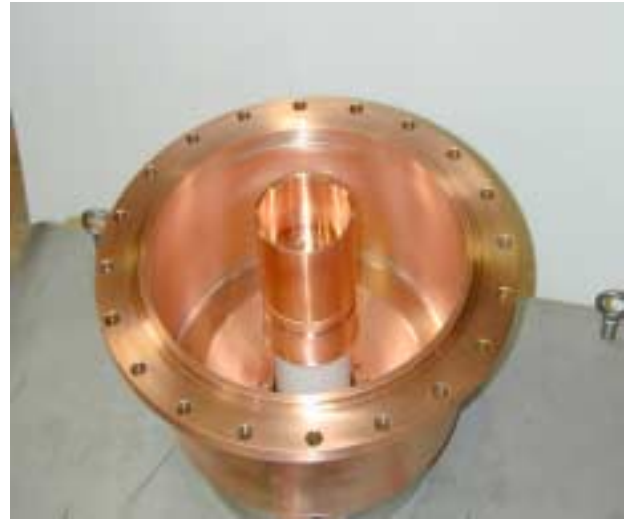

Figure 4: Tubular ceramic in a coaxial housing

The tubular feedthrough uses an Alumina ceramic from Jennings, which has an outer diameter of 2.38" and a length of 3.5" and is metalized at the edges. The ceramic assembly in a coaxial housing is shown in Fig. 4. Corona shields are incorporated in the design near the ceramic to metal joints. A SUPERFISH simulation is used to optimize the shape of the electrodes around the ceramic. Fig. 5 shows the electric field distribution in the ceramic and on the corona shields.

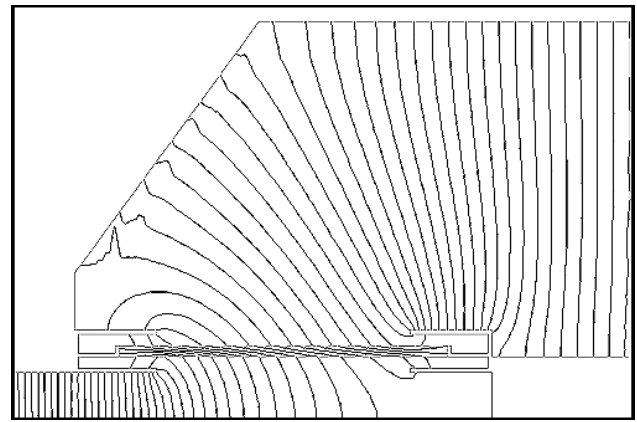

Figure 5: SUPERFISH simulation of e-fields in the tubular feedthrough

## 2.3 Insulating support for the coil

Since the coil is water cooled, it needs to be supported in order to reduce the vibration induced by the water flow. A polycarbonate material known as Lexan, is used as an insulator to support the coil. Tests shows that it can withstand high dc voltage and has low rf losses at the design frequency of the buncher. Unfortunately, it breaks down when rf is maintained for a while and catches fire and carbonizes. The mechanism of failure of Lexan under rf operation is not fully understood. This material is abandoned and a 2" diameter Teflon rod is used instead.

## 3 RF MEASUREMENTS

### 3.1 *Signal level*

The feedthroughs are assembled and connected to the coil, which is housed in the dust free box. The tubular feedthrough with the coil connected shows a Q of 1920 and a shunt impedance of 685 k$\Omega$ at 11.975 MHz. The capacitance of the feedthrough is measured to be approximately 27 pF. The disk feedthrough in parallel with the same coil shows a rather low Q value of 50. This implies that the ceramic is contaminated and further test is abandoned until the cause of such contamination is understood.

### 3.2 *Power*

The power test is done with the tubular feedthrough and the coil. The water cooling of the coil is extended to the ceramic-metal joint and the flow is 16 liters/minute. The vacuum of the test box is $2.10^{-7}$ Torr without rf applied. A 1 kW solid state amplifier is used for the test. A 4.5" diameter loop installed inside the coil at the short circuit end, is used to couple power. The loop can be turned to provide 50 $\Omega$ matching of the power amplifier and the resonant circuit. Under cw operation, 30 kV at the drift tube at 11.975 MHz is maintained for 6 hours without any breakdown or interruption. The maximum temperature on the ceramic is measured to be 38 degrees C. Measured values are shown in Table 2.

Table 2: Measured rf parameters of the prototype buncher

| Resonant Frequency | 11.975 MHz |
|---|---|
| Q, unloaded | 1920 |
| Rshunt | 685 k$\Omega$ |
| R/Q | 357 |
| Vtube | 30 kV |
| Pmeasured | 750 watts |
| Ptheoretical | 657 watts |
| Ceramic temperature | 38$^o$C |

The drift tube voltage is calibrated by measuring emitted x-rays. A glass window is provided in the test box for this purpose. Also, rf pick up probes are calibrated with the measured shunt impedance. Fig. 6 shows the measured tube voltage with input rf power varying from 800 watts to 1000 watts. This shows excellent agreement of the measured voltage with x-ray and rf pick up probes.

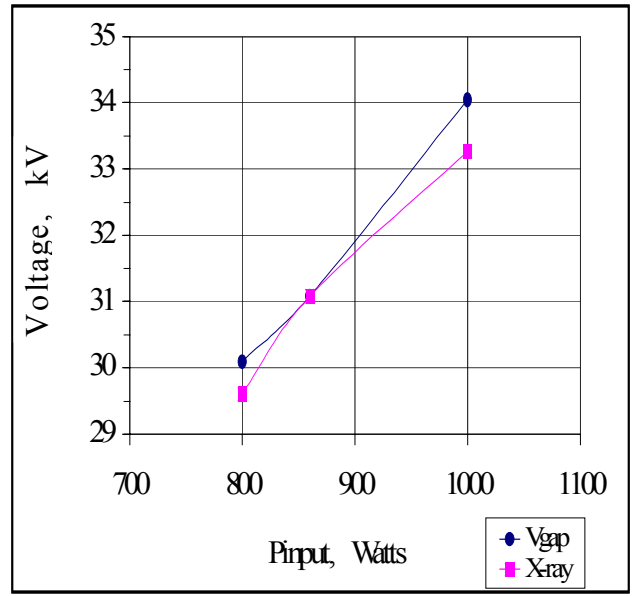

Figure 6: X-ray measurement of drift tube voltage

## 4 CONCLUSION

Since no commercial rf feedthrough is available which can withstand the 30 kV rf voltage at 11.8 MHz under cw mode, it has been decided to develop such a feedthrough at TRIUMF. The prototype tests have shown excellent high voltage performance of the tubular feedthrough. This will be used to design the final HEBT low-$\beta$ buncher. The 3 gap HEBT buncher requires two parallel circuits, consisting of two rf feedthroughs and two coils tuned to the same frequency. Hence, two fine tuners will be required for the operation of the buncher. A single coupling loop driven by a 2 kW power amplifier will be adequate.

## 5 ACKNOWLEDGMENTS

The authors like to thank Erk Jensen, CERN, Switzerland for providing the design of the CERN disk ceramic feedthrough. Thanks are due to Joseph Lu for making the coil, the box assembly and rf measurements, Al Wilson for the detail drawings of disk and tubular feedthroughs. Thanks are also due to Balwinder Waraich and Peter Harmer for providing technical assistance and Mindy Hapke for the photographs. We also wish to thank S. Arai, KEK, Japan and R.A Rimmer, LBL, USA for many helpful discussions.

## 6 REFERENCES

[1] R.E Laxdal, "Design Specification for ISAC HEBT", TRIUMF Design Note, TRI-DN-99-23
[2] R. Hohbach, "Discharge on ceramic windows and gaps in CERN PS cavities for 114 and 200 MHz", CERN/PS 93-60 (RF)